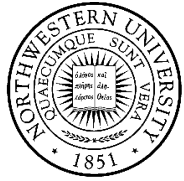

# NORTHWESTERN
## UNIVERSITY

Computer Science Department

Technical Report
NWU-CS-02-14
November 23, 2002

# A Tangible User Interface for Assessing Cognitive Mapping Ability

Ehud Sharlin, Benjamin Watson, Steve Sutphen, Lili Liu,
Robert Lederer, John Frazer

## Abstract

Wayfinding, the ability to recall the environment and navigate through it, is an essential cognitive skill relied upon almost every day in a person's life. A crucial component of wayfinding is the construction of cognitive maps, mental representations of the environments through which a person travels. Age, disease or injury can severely affect cognitive mapping, making assessment of this basic survival skill particularly important to clinicians and therapists. Cognitive mapping has also been the focus of decades of basic research by cognitive psychologists. Both communities have evolved a number of techniques for assessing cognitive mapping ability. We present the Cognitive Map Probe (CMP), a new computerized tool for assessment of cognitive mapping ability that increases consistency and promises improvements in flexibility, accessibility, sensitivity and control. The CMP uses a tangible user interface that affords spatial manipulation. We describe the design of the CMP, and find that it is sensitive to factors known to affect cognitive mapping performance in extensive experimental testing.

This research partially funded by an operating grant from the Canadian National Science and Engineering Research Council (NSERC).

# A Tangible User Interface
# For Assessing Cognitive Mapping Ability

Ehud Sharlin, Benjamin Watson[1], Steve Sutphen, Lili Liu[2], Robert Lederer[3], John Frazer[4]

Dept. Computing Science, University of Alberta
[1]Dept. Computer Science, Northwestern University
[2]Dept. Occupational Therapy, University of Alberta
[3]Dept. Art and Design, University of Alberta
[4]School of Design, Hong Kong Polytechnic University

**ABSTRACT**

Wayfinding, the ability to recall the environment and navigate through it, is an essential cognitive skill relied upon almost every day in a person's life. A crucial component of wayfinding is the construction of cognitive maps, mental representations of the environments through which a person travels. Age, disease or injury can severely affect cognitive mapping, making assessment of this basic survival skill particularly important to clinicians and therapists. Cognitive mapping has also been the focus of decades of basic research by cognitive psychologists. Both communities have evolved a number of techniques for assessing cognitive mapping ability. We present the `Cognitive Map Probe (CMP)`, a new computerized tool for assessment of cognitive mapping ability that increases consistency and promises improvements in flexibility, accessibility, sensitivity and control. The `CMP` uses a tangible user interface that affords spatial manipulation. We describe the design of the `CMP`, and find that it is sensitive to factors known to affect cognitive mapping performance in extensive experimental testing.

**Keywords**

Cognitive maps, wayfinding, cognitive assessment, neuro-psychological assessment, tangible user interfaces, constructional ability, spatial ability

**INTRODUCTION**

Almost every day, people find their way from home to any of a myriad of destinations, and then back again. Most take this skill for granted, but is an amazingly complex ability that has been the subject of decades of research by cognitive psychologists, who call it *wayfinding*. Injury or disease can so impair this ability that many become homebound, and for some unfortunate people, catastrophic failure of their wayfinding ability has lead to death from exposure. Thus medical researchers and clinicans also have a very strong interest in wayfinding.

A crucial component of wayfinding ability is cognitive mapping. A *cognitive map* is a mental representation of a person's environment, relied upon during wayfinding. Many techniques have been developed over the years for measuring and assessing this ability. Map drawing or placement is quite common, but is difficult to score consistently, wholly two-dimensional (2D) and necessarily quite abstract in representation. A few researchers have assessed cognitive mapping by asking patients or study participants to arrange 3D objects representing elements of their environment. While the reduced level of abstraction and more three-dimensional (3D) representation likely increases assessment sensitivity, previous implementations of this approach were quite unwieldy and still inconsistently scored.

To address these problems in assessment, we have designed the `Cognitive Map Probe (CMP)`, an automated tool for the measurement of cognitive mapping ability. The `CMP` makes use of the Segal model [12,13], a tabletop tangible user interface (TUI) originally designed for the input of architectural models. `CMP` users view a drivethrough of a neighborhood on a large screen perspective display, and then input their recollection of that neighborhood by arranging 3D building models on the Segal model's tabletop input surface. The `CMP` automatically records and scores each change the user makes to the model configuration. The `Cognitive Map Probe` is the first TUI for the assessment of cognitive mapping ability, combining the increased sensitivity of 3D input and affordances with the improved consistency, efficiency, flexibility and high-resolution data collection of computerization.

We begin the remainder of this paper with a review of what is known about cognitive maps, including their importance in everyday life and their measurement. A detailed description of the `CMP` follows, including comparisons to related TUIs. We conclude with a rigorous experimental examination of the sensitivity of the `CMP` to age and task

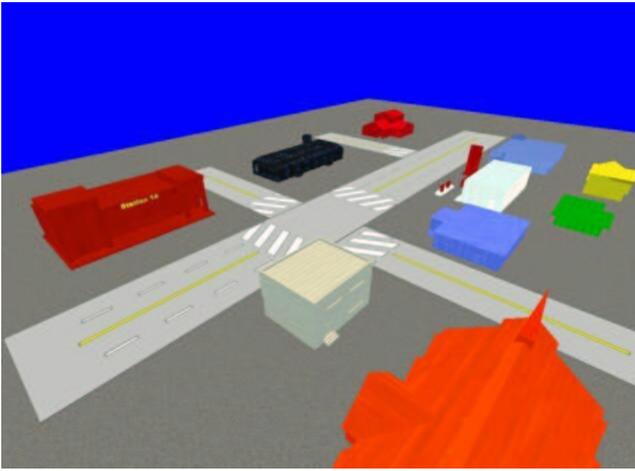 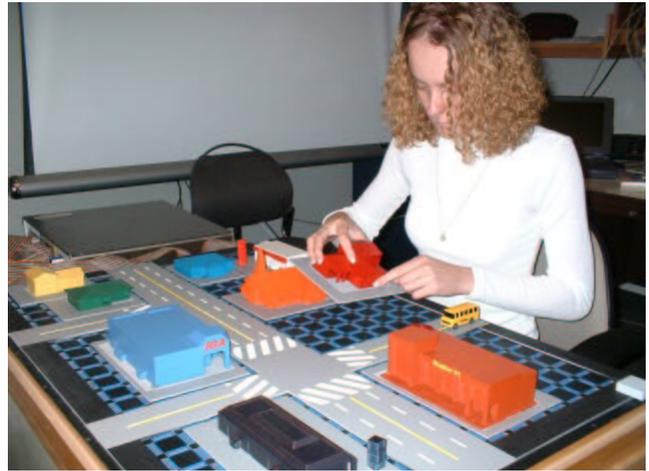

**Figure 1:** The virtual neighborhood as displayed by the CMP. This exocentric view was never used in experimentation, and is shown here only for illustration.

**Figure 2:** A participant manipulates the CMP's input board and building models.

difficulty, two factors that have a well-known relationship to cognitive mapping performance.

## MEASURING COGNITIVE MAPS

In his pioneering 1948 paper [25], Tolman argues that rats, like humans, have a mental representation of the world he called a cognitive map. These maps hold detailed spatial information that is collected, integrated and used while interacting with the environment. Tolman's work has led to the modern psychological definition of a cognitive map: *an overall mental image or representation of the space and layout of a setting* [1].

It is important to distinguish between the psychological concepts of wayfinding and of cognitive maps. Wayfinding refers to the overall process of reaching a destination [8], while cognitive maps underlie the wayfinding process and enable making and executing decisions about the environment [1,8].

Although the true nature of cognitive maps is not well understood, the most widely accepted theory of cognitive mapping is the Landmark-Routes-Survey (LRS) model [8]. The model divides our environmental understanding into three levels – landmark, route and survey – that can be integrated into a single comprehensive cognitive map [4,15,8].

Cognitive maps can often be imprecise. We tend to classify and cluster the massively detailed cognitive spatial information with which we are faced using simplifications, such as the gathering of objects and landmarks into hierarchies and regions. Cognitive maps also suffer from geometrical scaling and regularization problems [15].

Cognitive maps can be acquired through various means of interaction with an environment. Interaction can be classified as direct physical interaction, including walking down the street as well as a blind person tapping with a cane; or mediated through a variety of indirect means, including maps and virtual environments (VEs) [15,8].

Cognitive mapping ability is known to be affected by a number of cognitive- and task-related factors. Among these are age and task difficulty. Ability is also related to various forms of dementia such as Alzheimer's disease (AD), to such an extent that the missing person waiting period is waived for diagnosed dementia patients, who have died from exposure when they become lost and disoriented. Some have proposed assessment of cognitive mapping ability as a form of AD diagnosis [18]

### Cognitive Maps in Virtual Environments

Many researchers have experimented with VEs as wayfinding training tools [6,17,8]. The technology that supports these applications is diverse and ranges from low-end desktop PCs to CAVEs employing treadmills as travel interfaces [5,6,17]. While it seems obvious that the highest levels of immersion should be employed for wayfinding training, one study has shown that training based on an expensive stereoscopic HMD was no more effective than training that used a simple digital projector [19].

A concern that overshadows VE-based wayfinding trainers is the problem of training transfer, that is determining whether the cognitive map acquired in the VE is useful in the physical world. Currently there is no clear-cut answer to these questions [5,17]. It has also been demonstrated that VE wayfinding training might actually hinder the development of survey knowledge. At the same time, it might be a useful way of assessing real world wayfinding ability [7].

### Techniques for Probing Cognitive Maps

There are several techniques for assessing cognitive mapping ability. Verbal techniques simply ask a person to describe the environment. These techniques suffer from the subjective nature of the reported information and from natural variability in communicative ability. However, verbal techniques can achieve deep insight into cognitive mapping through use of verbs of motion rather than just dry description of physical locations [2].

The bearing and distance technique [3,4,6,14,17] places a person at a certain location in the environment, and asks him or her to point to other objects in the environment and estimate the distances to them. These inter-object distances and directions are then compared to the distances in the original environment. The bearing and distance technique is easy to implement, but the technique suffers from scale problems and may not be very sensitive to survey knowledge enabling generation of new paths through the environment [5].

Map drawing [5] or placement [3,14,20] techniques ask a person to describe his or her cognitive map through sketching or placing prepared objects. Drawing techniques are sensitive to variation in sketching ability. In work of particular relevance here, 3D objects or models have been used in map placement. For example, Piaget used a "Model Village" with cardboard models of a church, houses and trees to help children input cognitive maps [20].

Functional assessment techniques position a person back in a previously studied spatial environment and assesses the person's ability to perform a novel navigation [14]. This technique can provide excellent insight into the user's survey knowledge, but requires much assessor time and can raise a psychological Heisenberg-like principle as the ability measured is altered by measurement [5].

The use of computers in cognitive map assessment is very limited. The first use dates to the late 70s when Baird designed a computerized map placement assessment technique. The technique used a now obsolete computer interface for inputting building locations on a 13 x 13 matrix displayed on a monitor [3]. Computerized tools have also been used for automatic collection of bearing estimations (see for example [4]).

**THE COGNITIVE MAP PROBE**

The CMP is an automated system for assessment of cognitive mapping ability. During the first phase of each trial, the participants view a virtual neighborhood displayed with a digital projector (fig. 1). Viewing can be passive, similar to riding in a bus; or active and more akin to participants driving the bus. Viewing can also be egocentric, with participants seeing a street level view; or exocentric, with participants seeing a bird's eye view.

In the trial's second phase, participants move to a 2D input surface and construct their cognitive map of the neighborhood they have just seen (fig. 2). This input is accomplished by arranging physical, 3D models of buildings on the Segal model's 2D board. When participants place or remove buildings from the board, the system records the building ID, its 2D location and the time of the event. During placement the system also records the building's orientation. When participants are satisfied that the constructed configuration accurately represents their cognitive map, they signal the assessor who advances the system to the next trial.

**Hardware and Software**

We printed the CMP's user interface by creating 10 virtual building models in a software package, and then outputting them in 3D using rapid prototyping technology. The resulting polyester objects are quite sturdy and mounted on flat bases, under which is a single connector for the Segal model's board. Aligning the base with the board's slots aligns the connector to its matching slot and eases insertion of the model. All the models are of similar scale and can be arranged easily with two hands. The models were spray painted in single primary colors for easy viewing by the elderly, but important details such as store signs were hand painted in contrasting colors. The models are quite detailed in shape, and include doors, windows, and even the patterns of wood siding. We also attached a simple street pattern to the board (one four-way and one "T" intersection, see figs. 1 and 2); this street pattern was never removed during assessment. All 10 models and the street pattern can fit onto the board at the same time.

The virtual versions of these physical models also populate the virtual neighborhoods shown to participants in the first phase of each assessment trial. Thus buildings in the displayed virtual neighborhoods match the physical models used for tangible interaction exactly in shape and nearly exactly in color.

The Segal model is a pioneering TUI named in memory of architect and advocate of home self-design Walter Segal. John Frazer and his colleagues built the Segal model [12,13] in collaboration with Segal to support his work. The device was designed to enable direct, tangible interaction with architectural floor plans and their components, such as walls, doors, windows, plumbing fixtures and furniture. It is a 102cm x 71cm board covered with an array of 768 edge connector slots arranged in 24 columns of 16 vertical slots and 16 rows of 24 horizontal slots. Each slot has contacts enabling recognition of 127 different connector types, after accounting for symmetries in orientation. Architectural components were represented by physical 3D models, with each type of component coupled to a unique connector type. Since our application required tangible, tabletop interaction very similar to that supported by the Segal model, we converted it for our use.

After assessment, the CMP analyzes the data it has collected to score the participant's performance. As we discussed above, there are a number of ways in which cognitive maps may be scored. All involve comparisons of the actual map $M$ to the participant's cognitive map $C$. Measures that disregard position and treat $M$ and $C$ only as sets of buildings are:

$$number = 1 - abs(|M|-|C|)/|M|$$
$$difference = 1 - (|M-C|+|C-M|)/(|M|+|C|)$$

Measures that compare position only within the set of intersecting buildings $M \cap C$ include:

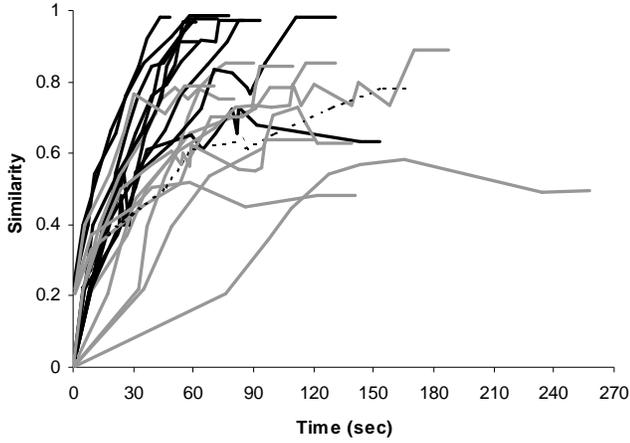

**Figure 3:** Similarity vs. time for all participants in the trial with 8 buildings. Young are shown in black, elderly in gray. The single mild AD participant is shown as a dotted line.

$$distance = 1 - \Sigma_i \ (dist(M_i,C_i)/d_{max})/m_{max}$$
$$orient = 1 - \Sigma_i \ (odiff(M_i,C_i)/180)/m_{max}$$
$$interbuilding = 1 - \Sigma_i\Sigma_j \ (abs(D_{Mij}-D_{Cij})/d_{max})/m_{max}^2$$

where all sums range over the set $M \cap C$, $dist$ is the Euclidian distance function, $odiff$ is the angular difference in degrees between the orientation of two buildings, $d_{max}$ is the length of the CMP board diagonal, $m_{max}$ is the maximum of $|M|$ in the entire assessment, and $D_M$ and $D_C$ are square matrices in which the entries are $dist(M_i,M_j)$ and $dist(C_i,C_j)$, respectively, with i and j again ranging over the set $M \cap C$. Finally, the CMP forms a composite measure that includes both set and position error:

$$similarity = difference \times distance \times orient$$

Recall that the CMP also records the time of each action on the board. This allows us to add *totalTime*, the time it takes to complete one assessment trial, to our suite of measures. We can also probe the progress participants make during the assessment by comparing our measures to the current time. Figure 3 graphs *similarity* vs. time for all participants in one assessment trial. We construct the additional measure *dSim* by finding the differences between consecutive measurements of *similarity* divided by the time elapsed between those measurements, and averaging the resulting "local slopes" over all such pairs in an assessment trial.

### Related Systems

According to Ishii and Ulmar TUIs are *devices that give physical form to digital information, employing physical artifacts as representations and controls of the computational data* [26]. In [24] spatial TUIs are defined as *interface devices that use physical objects as means of inputting shape, space and structure into the virtual domain*. Using the CMP, we wanted to realize the potential of TUIs for spatial input in a practical cognitive assessment application, which called for tabletop interaction.

Several tabletop TUIs have been developed employing various technologies (for a more general review of TUIs, see [26]). Vision-based TUIs include BUILD-IT [11], URP [27] and Illuminating Light [28]. All use passive vision tracking for meditating interaction using simple tangible objects. Illuminating Clay [21] uses active vision tracking to enable users to alter topography tangibly in a graphical landscape analysis application.

Radio frequency identification (RFID) systems include DataTiles [22] and the Senseboard [16]. Objects are tagged with RFIDs to enable tabletop interaction for organizing information (Senseboard) and as modular construction units (DataTiles).

Although not TUIs by definition, DiamondTouch [9] and PingPongPlus [29] use a tabletop interaction metaphor. PingPongPlus tracks a ping pong ball using sonic tracking, while DiamondTouch uses a touch sensitive tabletop interface, which is based on closing a capacitively coupled circuit through the user's finger.

All these interfaces also output information back to the tabletop using front projection.

Bricks [10] and RUGAMS [23] use a tabletop metaphor and magnetic trackers to enable interaction. Bricks also makes use of a pressure-sensitive tablet.

### System Strengths

The CMP offers the following advantages over existing methods for assessing cognitive mapping skill:

*Sensitivity*. The CMP monitors participant progress (or lack thereof) throughout map construction. In contrast, existing methods assess cognitive mapping only when the map is complete. In addition, the CMP's 3D tangible interface allows a much more direct translation of cognitive maps into physical representations, with fully detailed buildings viewable in perspective from all sides, much as they are during travel through the represented neighborhoods themselves. Commonly used 2D assessment methods offer only highly abstracted 2D projections of the represented environment and its buildings. Ultimately, it should be possible to add adaptivity to the CMP, focusing more quickly and completely on the limits of participant ability, and improving sensitivity further.

*Accessibility*. Many of the populations commonly given cognitive mapping assessments face cognitive, visual or motor challenges. Unlike traditional 2D assessment techniques, the CMP uses an interface that is intuitive, easy to see, and simple to manipulate. This proved invaluable during our work with the elderly.

*Consistency*. If an assessment is to have meaning outside of its original context, it must be performed consistently and reliably by all assessors. Existing 2D assessments are consistent, but achieving this consistency requires that the assessments be fairly simple to perform, reducing assessment sensitivity. Because it is automated, the CMP achieves the highest level of consistency while at the same time improving sensitivity with complex tasks and very frequent measurement of the participant.

*Control.* The CMP's virtual neighborhood display will always be simpler than real world stimuli. On the other hand, virtual display offers an amazing degree of control in assessment. Climates can be changed, landmarks rotated or removed, buildings located incorrectly by the participant can be displayed translucently on top of correctly located buildings, and neighborhoods can be viewed from positions in midair – effects extremely difficult if not impossible to achieve in the real world.

## ASSESSING THE COGNITIVE MAP PROBE

How sensitive is the CMP to well-known cognitive factors in practice? In this section we describe the experiment we performed to find answers to this question. We also describe what we learned about the accessibility and consistency of the CMP as we put it through its paces.

### Methodology

The CMP was designed to support a wide range of cognitive mapping tasks. In this experiment, we sampled this range by varying the *number of buildings* in the virtual neighborhood we asked participants to recreate.

We expected that cognitive mapping performance would worsen by all measures as the *number of buildings* in the mapped environment increased. We also anticipated that performance among our elderly participants would be worse than the performance of our young participants, reflecting the natural effects of *age* on cognitive ability.

*Participants*

Our experiment had 20 participants. They were recruited from on and off campus, and ranged in *age* from 25 to 81. There were ten young participants under 55 years old, and ten elderly participants 55 aged years or more. Both groups were balanced in gender. As a preliminary study, we also worked with one additional participant who had been diagnosed with mild AD. This single participant was not included in any experimental results except where noted.

*Design*

Each participant performed 7 recorded trials. The virtual neighborhoods in each of the remaining 7 trials were composed of a unique number of buildings. All participants viewed the same virtual neighborhoods in the same order, with the *number of buildings* increasing from 2 to 8. Neighborhoods were ordered in this fashion so that thresholds in participant cognitive ability could be quickly identified without subjecting participants to unnecessary confusion or frustration. (3 participants not included elsewhere in this discussion or results were in fact unable to complete all 10 assessment trials).

*Apparatus and Procedure*

All experiments were conducted according to a strict written protocol, and with a script read out loud to each participant. In the script, the participant was introduced to the CMP, the experiment, and its purpose, and then read an information letter. The participant was told that he or she might stop the experiment at any time, and asked to sign a consent form. The participant was interviewed quickly, answering questions concerning age, education and occupation. Participant anonymity was always preserved.

Accuracy was emphasized over precision in instruction, with participants asked to be as precise as possible, but reminded that the CMP was recording the speed of their actions. Participants were told that there was no time limit, that they may decide when they had finished each task, but that they should do the best they could in reconstructing each neighborhood.

The assessor guided participants through three initial practice trials to train them in the use of the CMP. All practice trials used simple two buildings neighborhoods. In the first trial, the assessor introduced the CMP board and its models, as well as a "bus ride" metaphor for the largely passive, egocentric viewings participants would have of virtual neighborhoods. The assessor then talked participants through a viewing of the virtual neighborhood that corresponded to the map already on the board. The assessor made certain that the participant understood this virtual-physical correspondence, and demonstrated that the passive viewing might be paused at will for a panoramic viewing (see below). In the second trial, the assessor introduced board interaction to the participant by asking the participant to identify a slight change to the virtual neighborhood during a new virtual tour. The assessor then turned off the virtual neighborhood display and asked the participant to adjust the CMP board to match this changed virtual neighborhood. In the third trial, the assessor confirmed that participants completely understood typical interaction by having participants view a completely new virtual neighborhood, and asking them to recreate it on the CMP board, again after the virtual neighborhood display was turned off.

During the first phase of a recorded trial, participants viewed a virtual neighborhood from a passive, egocentric perspective, moving through the neighborhood at street level. A compass in the ground plane indicated which direction was north. All participants moved along the same path. Participants could optionally halt their motion at any time and rotate slowly through 360 degrees for a panoramic viewing before continuing along the viewing path. The virtual neighborhood display was then turned off and participants moved into the trial's second phase, during which they interacted with the board and attempted to reconstruct the neighborhood they had just viewed from memory. A physical pointer similar to the compass seen in the first phase indicated which direction was north. Participants never received any feedback or comments about their performance from the CMP or the assessor. Participants required 1 ½ hours on average to complete the full set of 3 practice and 7 recorded trials, as well as a short post-assessment interview.

### Results

Since the Segal model is a historic interface, we fully expected some noise in data collection. However, the CMP

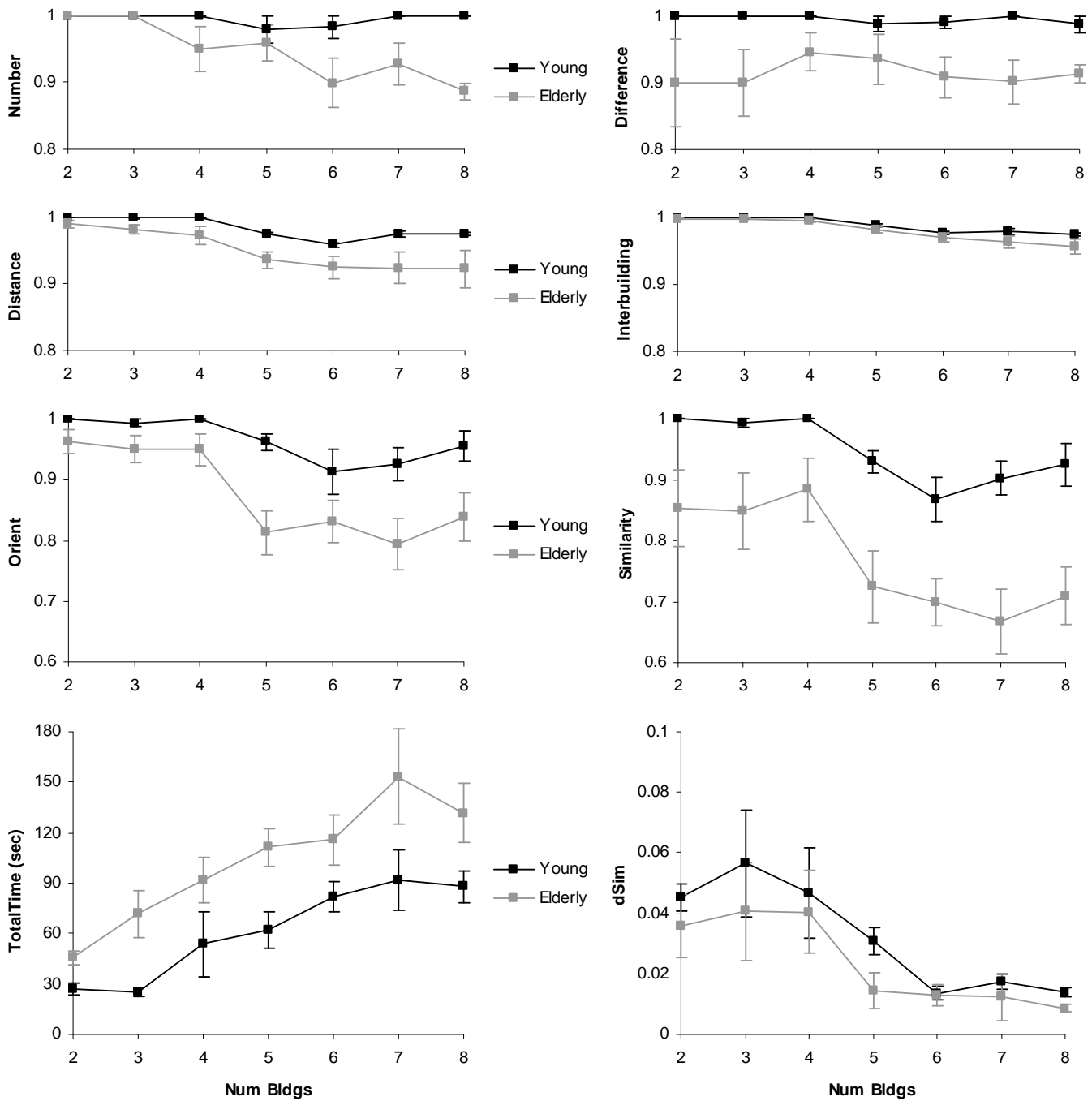

**Figure 4:** Sensitivity study results, showing experimental means and standard errors. AD participant excluded.

performed relatively well. Most importantly, no participant was forced to repeat a trial. The CMP also made no errors when reporting location. Nevertheless, there were errors when reporting the identity of the buildings attached or detached from the board. The only such errors that could not be corrected automatically were unidentified buildings, and misidentified buildings. Unidentified buildings made up 18% of all actions on the board and were corrected interactively by the assessor during the trial. Misidentified buildings made up less than 2% of all actions (21 actions total), but had to be corrected after assessment by manually matching CMP data to video recordings of the assessment. Though annoying, both types of errors occurred at rates quite manageable for our purposes and we are confident that a more polished implementation, possibly using different base technology, could eliminate most if not all of these problems.

Figure 4 presents our experimental results by all dependent measures. We analyzed these results with one ANOVA for each dependent measure. Each such analysis was two-way, (2 *age* x 7 *num buildings*), with *age* a between subjects factor, and *num buildings* a within subjects factor. Results of these analyses are presented in Table 1.

**Table 1:** Results of two way ANOVAs in sensitivity study. AD participant excluded.

| Indep Meas | Depend Meas | ANOVA |
|---|---|---|
| age | totalTime | $F(1,18)=9.242, p=.007$ |
| age | number | $F(1,18)=14.797, p=.001$ |
| age | difference | $F(1,18)=14.928, p=.001$ |
| age | orientation | $F(1,18)=15.501, p=.001$ |
| age | distance | $F(1,18)=6.250, p=.022$ |
| age | interbuilding | $F(1,18)=4.782, p=.042$ |
| age | similarity | $F(1,18)=18.844, p<.0005$ |
| # bldgs | totalTime | $F(6,108)=15.432, p<.0005$ |
| # bldgs | number | $F(6,108)=3.400, p=.004$ |
| # bldgs | orientation | $F(6,108)=9.823, p<.0005$ |
| # bldgs | distance | $F(6,108)=12.290, p<.0005$ |
| # bldgs | interbuilding | $F(6,108)=23.800, p<.0005$ |
| # bldgs | similarity | $F(6,108)=9.333, p<.0005$ |
| # bldgs | dSim | $F(6,108)=7.567, p<.0005$ |
| age x # bldgs | number | $F(6,108)=2.884, p=.012$ |

The CMP responded very much in line with our expectations to the cognitive factor *age* and the task factor *num buildings*. In the seven measures that responded significantly to *age*, the elderly were uniformly worse in cognitive mapping performance. In the seven measures that responded significantly to *num buildings*, response was more complex, with measures worsening initially as the number of buildings increases, then reaching a plateau or even improving slightly as the number of buildings reached maximum. It may be that when the number of buildings was high, the additional location constraints imposed by the physical street pattern on the board limited the number of possible configurations and made cognitive mapping easier. Alternatively or additionally, since trials with larger neighborhoods were always encountered later in the assessment, participants may simply have been more practiced by the time these larger neighborhoods were encountered.

Only *dSim* failed to respond significantly to *age*. Trends in the data indicated that rates of mapping progress for the young might become larger than rates for the elderly, were experimental sample size increased. Similarly, only *difference* did not vary significantly as *num buildings* changed. Here the null hypothesis – that the normalized set difference is simply not sensitive to the size of the map participants are attempting to reproduce – likely provides the best explanation of this result. However, an interesting reflective symmetry in the young and elderly curves (see fig. 4) may indicate opposite and canceling responses to the number of buildings.

The effects of *age* and *num buildings* interacted only in the *number* measure. While *num buildings* had little effect on the young, the mapping performance of the elderly dropped significantly by this measure as the number of buildings increased. This is likely due to an age-based difference in recall.

**Discussion**

In this section we review the broader implications of our results for the CMP. We begin, however, by noting again that because of our need to find the cognitive thresholds of our participants quickly, we ordered experimental trials so that the *num buildings* factor increased steadily. Because of this pointed lack of counterbalancing or randomization in *num buildings*, practice effects are confounded with the observed effects of *num buildings*.

*Confirmations*

Our experimentation confirms that the CMP is sensitive to factors known to affect cognitive mapping performance. As expected, the bulk of our results indicate that the elderly perform worse at cognitive mapping than the young. Increasing the size of the map being reproduced can also worsen mapping performance.

We were also pleased with the match of the CMP interface to the mapping task, and its accessibility to the elderly population. Almost all of our participants were able to complete all trials – and most reported they had fun doing so. This was true whether participants were university students or World War II veterans.

Our results are very preliminary, but we were also gratified to see that our single AD participant was among the worst performers, tentatively indicating possible use of the CMP for palliative care of persons with AD. Much more research is required before this application is realized.

*Surprises*

We expected that assessment performance would worsen as *num buildings* increased. Instead, *num buildings* had a much more complex impact. While confounding practice effects certainly had an influence on this result, the initial *decrease* in mapping performance as the number of buildings increases (the opposite of a practice effect) leads us to believe that the constraints provided by our tangible street pattern played a larger role. This suggests that mapping difficulty might be controlled in future experiments by varying proportion of the map used for street cues.

We did not expect the *age* x *num buildings* interaction we saw in our results. It would be interesting to see if performance in the *number* measure also declines for the young as the number of buildings increases further.

*Implications*

While our results indicate great promise for technologies like the CMP, there is much work that remains if its assessment paradigm is to become common in clinical and research settings. First, the measurement sensitivity and reliability of CMP-like tools must be probed further, with comparisons made to existing assessment techniques, and typical score distributions found so that unusual assessment results might quickly be recognized. Second, tangible and tabletop interaction must become cheaper and more reliable, so that newer versions of the CMP will be more cost effective.

We believe that future innovation in tangible UIs must, like the CMP, be closely tied to target applications. Such close relationships will allow researchers to isolate those components of the tangible interface that are both strong and weak, in concrete application terms. For example, we have already built a parallel WIMP-based cognitive mapping assessment tool, and plan to compare its sensitivity to the CMP. Pushing TUIs closer to applications should also eventually bring them into production, which will make them more cost-effective and widespread.

**CONCLUSION**

In this paper we have presented the CMP, a tangible user interface for the assessment of cognitive mapping ability. In experimentation, the CMP proved to be sensitive to factors known to affect cognitive mapping ability.

Our work on the CMP will continue. There are many interesting opportunities for improving its sensitivity. For example the CMP could be used iteratively, with visual feedback given to the participant about the accuracy of the currently reproduced map, enabling the participant to attempt to correct their map. Active or exocentric viewing modes might be explored. The detailed histories of map building compiled by the CMP might be analyzed to find the decision trees formed by participants. Ultimately, the CMP might also prove useful therapeutic applications.